\documentstyle[12pt]{article}
\begin{document}
\begin{center}
\Large\tt\bf{Gravitational field of global monopoles in Einstein-Cartan gravity}
\end{center}
\vspace{2cm}
\begin{center}
by L.C.Garcia de Andrade\footnote{Depto. de  F\'{\i}sica  Te\'{o}rica, 
Instituto de F\'{\i}sica - UERJ, Rio de janeiro,20550, Brasil} 
\vspace{3cm}
\end{center}
\begin{abstract}
The gravitational weak field of a global monopole in the 
Einstein-Cartan theory of gravity is investigated.To obtain this solution 
we assume that Cartan torsion takes the form of the Newtonian gravitational 
potential.From the geodesics it is possible to show that the torsionic 
monopole produces a repulsive gravitational force. 
\end{abstract}
\newpage
Recently topological defects \cite{1} have been considered in the context of alternative 
theories of gravity such as Brans-Dicke theory of gravity \cite{2} and Einstein-Cartan theory 
of gravity \cite{3}.In particular Barros and Romero \cite{2} extended Barriola-Vilenkin \cite{4}
solution for a global monopole in General Relativity.In all these papers gravitational fields 
representing repulsive forces were found.In this letter we show that it is possible to find a 
global monopole in Einstein-Cartan gravity in linear approximation regime by considering the 
ansatz that torsion has radial symmetry and that it has the form of a Newtonian gravitational 
potential.This ansatz not only simplifies a lot our computations but also is physically 
acceptable.Let us consider the general spherically symmetric metric in the form   
\begin{equation}
ds^{2}=e^{N(r)}dt^{2}-e^{L(r)}dr^{2}-r^{2}(d{\theta}^{2}+sin^{2}{\theta}d{\phi}^{2})
\label{1}
\end{equation}
The energy-stress tensor of the static global monopole (outside the core) is
approximatly given by
\begin{equation}
T^{i}_{j}=diag\frac{{\eta}^{2}}{r^{2}}(1,1,0,0)
\label{2}
\end{equation}
where ${\eta}$ is the coupling constant for the monopole.We also consider the Einstein-Cartan 
field equation in the quasi-Einsteinian form where only the stress-energy tensor receives a 
contribution from torsion and is given by
\begin{equation}
{T_{i}^{k}}^{torsion}=S_{iml}S^{kml}-{\delta}_{i}^{k}{S_{0}}^{2}
\label{3}
\end{equation}
where torsion is given by the only non-vanishing component $S_{0}=S_{230}$ and the torsion 
tensor is taken as totally skew as for Dirac particles.Let us now consider the Einstein tensor 
for the metric (\ref{1}) in the linear approximation where we take the metric components as
\begin{equation}
e^{L}=(1+f(r))
\label{4}
\end{equation}
and
\begin{equation}
e^{N}=(1+g(r))
\label{5}
\end{equation}
Substitution of these last two expressions into the usual spherically symmetric components of 
the Einstein-Riemannian tensor we obtain the following expressions
\begin{equation}
G^{0}_{0}=-(1+f)^{-1}(\frac{f'}{(1+f)r}-\frac{1}{r^{2}})-\frac{1}{r^{2}}=\frac{{\eta}^{2}}{r^{2}}-{S_{0}}^{2}
\label{6}
\end{equation}
and
\begin{equation}
G^{1}_{1}=(1+g)^{-1}(\frac{g'}{(1+g)r}-\frac{1}{r^{2}})-\frac{1}{r^{2}}=\frac{{\eta}^{2}}{r^{2}}-{S_{0}}^{2}
\label{7}
\end{equation}
and 
\begin{equation}
G^{2}_{2}=G^{3}_{3}=(1+g)^{-1}(\frac{g"}{(1+g)}-\frac{1}{2r}(\frac{g'}{1+g}-\frac{f'}{1+f}))
\label{8}
\end{equation}
From equations (\ref{6}) and (\ref{7}) we immediatly notice that the ansatz ${S_{0}}^{2}=\frac{{\eta}^{2}}{r^{2}}$ 
simplifies a lot our computations.From the resultant equations one obtains like in the 
Schwarzschild case of General Relativity that $(1+g)=(1+f)^{-1}$ and some simple algebra leads 
to the differential equation
\begin{equation}
g"+\frac{f'}{2r}=-\frac{{\eta}^{2}}{r^{2}}
\label{9}
\end{equation}
From the first equation for $f$ we obtain another differential equation given by
\begin{equation}
f'+\frac{f}{r^{2}}=0
\label{10}
\end{equation}
Equation (\ref{10}) yields
\begin{equation}
f=e^{cr}
\label{11}
\end{equation}
Substitution of (\ref{11}) into (\ref{10}) yields
\begin{equation}
g"-\frac{ce^{cr}}{2r^{3}}=-\frac{{\eta}^{2}}{r^{2}}
\label{12}
\end{equation}
By taking the approximation where terms like $r^{-3}$ may be dropped we 
obtain the following final result
\begin{equation}
g_{00}=1-{\eta}^{2}r-{\eta}^{2}lnr
\label{13}
\end{equation}
which can also be expressed in terms of torsion as 
\begin{equation}
g_{00}=1-{S_{0}}^{2}r^{3}-{S_{0}}^{2}r^{2}lnr
\label{14}
\end{equation}
The Bariola-Vilenkin limit of a Einsteinian global monopole is 
obtained from (\ref{13}) by expanding the logarithm function $lnr$.
The gravitational nature of the field is obtained by considering the 
geodesics equation in linear approximation
\begin{equation}
\frac{d^{2}x^{i}}{dt^{2}}=-\frac{{\partial}h_{00}}{{\partial}x^{i}}
\label{15}
\end{equation}
where the perturbation of the metric is given explicitly.From the geodesics 
thus we may conclude that the gravitational force is repulsive.An exact 
global monopole solution of Einstein-Cartan field equations is now under 
investigation.
\section*{Acknowledgements}
I would like to thank Carl Brans and A.Wang for helpful discussions on the 
subject of this paper.Financial support of UERJ (FAPERJ) is gratefully 
acknowledged.


\begin{thebibliography}{4}
\bibitem{1}A.Vilenkin and P.S.Shellard.Cosmic Strings and Other Topological Defects(1995)Cambridge 
University Press.
\bibitem{2}A.Barros and C.Romero,Global monopoles in Brans-Dicke theory of Gravity(1985)gr-qc.
\bibitem{3}L.C.Garcia de Andrade,J.Math.Phys.(1998).
\bibitem{4}Barriola and A.Vilenkin,Phys.Rev.Lett,(1977).
\end{thebibliography}
\end{document}